\documentclass[pre,twocolumn,floatfix]{revtex4}

\usepackage{amssymb,amsmath}

\usepackage{graphicx}

\usepackage{subfigure}

\begin{document}

\title{A Symmetric Free Energy Based Multi-Component Lattice Boltzmann
Method} 

\author{Qun Li}
\author{A.J. Wagner}

\affiliation{Department of Physics, North Dakota State University,
Fargo, ND 58105}

\begin{abstract}
We present a lattice Boltzmann algorithm based on an underlying free
energy that allows the simulation of the dynamics of a multicomponent
system with an arbitrary number of components.  The thermodynamic
properties, such as the chemical potential of each component and the
pressure of the overall system, are incorporated in the model.  We
derived a symmetrical convection diffusion equation for each component
as well as the Navier Stokes equation and continuity equation for the
overall system.  The algorithm was verified through simulations of
binary and ternary systems. The equilibrium concentrations of
components of binary and ternary systems simulated with our algorithm
agree well with theoretical expectations.
\end{abstract}

\maketitle

\section{Introduction}
Multicomponent systems are of great theoretical and practical
importance. An example of an important ternary system, that inspired
the current paper, is the formation of polymer membranes through
immersion precipitation \cite{akthakul}. In this process a
polymer-solvent mixture is brought in contact with a non-solvent. As
the non-solvent diffuses into the mixture, the mixture
phase-separates, leaving behind a complex polymer morphology which
depends strongly on the processing conditions. The dependence of the
morphology on the parameters of the system is as yet poorly
understood. Preliminary lattice Boltzmann simulations of this system
exist \cite{akthakul}. However, this work did not recover the correct
drift diffusion equation. A general fully consistent lattice Boltzmann
algorithm with an underlying free energy to simulate multicomponent
systems is still lacking. This paper strives to bring us a step nearer
to achieving this goal.

There are several previous lattice Boltzmann methods for the
simulation of multi-component systems. There are three main roots for
these approaches. There are those derived from the Rothmann-Keller
approach \cite{RK,Reis} that attempt to maximally phase-separate the
different components. A second approach by Shan and Chen is based on
mimicking the microscopic interactions \cite{Shan1,Shan2,shanx} and a
third approach after Swift, Orlandini and Yeomans \cite{osborn,orlandini} is
based on an underlying free energy. All of these have different
challenges. Since we are interested in the thermodynamics of
phase-separation we find it convenient to work with a method based on
a free energy. This allows us to easily identify the chemical
potentials of the components. This is convenient since the gradients
of the chemical potentials drive the phase separation as well as the
subsequent phase-ordering.
 
The challenge for the LB simulation of a multicomponent system lies in
the fact that momentum conservation is only valid for the overall
system but not for each component separately, and diffusion occurs in
the components.  For a binary system of components $A$ and $B$ with
densities $\rho^A$ and $\rho^B$, the simulation usually traces the evolution
of the total density $\rho^A + \rho^B$ and the density difference $\rho^A -
\rho^B$ \cite{orlandini}.  Although this scheme is successful in the
simulation of a binary system \cite{orlandini,wagner2001}, its
generalization for the LB simulations of systems with an arbitrary
number of components is asymmetric.  For instance, to simulate a
ternary system of components $A$, $B$, and $C$ with densities $\rho^A$,
$\rho^B$ and $\rho^C$, the total density of the system, $\rho^A + \rho^B + \rho^C$,
should be traced, and the other two densities to be traced may be
chosen as, e.g., $\rho^B$ and $\rho^A - \rho^C$ \cite{lamura}.  This approach
is likely to be asymmetric because the three components are treated
differently as is the case of Lamura's model \cite{lamura}.  If an LB
method is not symmetric, it will lose generality an will only be
adequate for special applications.  In this paper, we established a
multicomponent lattice Boltzmann method based on an underlying free
energy that is manifestly symmetric.
 
\section{Macroscopic  Equations  for Multicomponent System}
The equation of motion for a multicomponent system are given by the
continuity and Navier-Stokes equations for the overall system and a
drift diffusion equation for each component separately. The continuity
equation is given by
\begin{equation} 
\partial_t \rho + \nabla \cdot  {\bf J}  = 0 , \label{continuityEQ} \\ 
\end{equation} 
where $ \rho $ is the mass density of the fluid, ${\bf J}$ is the mass
flux which is given by ${\bf J} \equiv \rho \, {\bf u}$, and ${\bf u}$ is
the macroscopic velocity of the fluid.  The Navier-Stokes equation
describes the conservation of momentum:
\begin{equation} 
\partial_t(\rho u_\alpha) + \partial_\beta(\rho u_\alpha u_\beta) 
= -\partial_\beta P_{\alpha \beta}+ \partial_\beta \sigma_{\alpha \beta} + \rho F_{\alpha} , \label{NSEeq} 
\end{equation} 
where $P_{\alpha\beta}$ and $\sigma_{\alpha \beta}$ are the pressure and viscous stress
tensors respectively, ${F_\alpha}$ is the component
$\alpha$ of an external force on a unit mass in a unit volume, and the
Einstein summation convention is used.  For Newtonian fluids, the
viscous stress tensor is given by
\begin{equation} 
\sigma_{\alpha \beta} = \eta \left( \partial_\beta u_\alpha + \partial_\alpha u_\beta - \frac{d}{2}\delta_{\alpha \beta} 
 \nabla \cdot {\mathbf u} \right) + \mu_B \delta_{\alpha \beta} \nabla \cdot {\mathbf u},
\label{stresstensor}
\end{equation} 
where $\eta$ is the shear viscosity, and $\mu_B$ is bulk viscosity, and $d$
is the spacial dimension of the system.

Free energy, chemical potential, and pressure are key thermodynamic
concepts to understand the phase behavior of a system.  The chemical
potential of each component can be obtained by a functional derivative
as
\begin{equation} 
\mu^{\sigma} = \frac{\delta {\mathcal F}}{\delta n^\sigma}  \label{chemicalpotential},
\end{equation}   
where $\mu^\sigma$ is the chemical potential of component $\sigma$; $n^\sigma$ is the
number density of component $\sigma$; and $\mathcal F$ is the total free
energy of the system.

The pressure in a bulk phase in equilibrium is given by
\begin{equation} 
p = \sum_\sigma n^\sigma \mu^\sigma - \psi .  \label{ppp}
\end{equation}  
The pressure tensor is determined by two constraints: $ P_{\alpha \beta} = p
\delta_{\alpha \beta}$ in the bulk and $\Delta P_{\alpha \beta} = \sum_\sigma n^\sigma \nabla \mu^\sigma$ everywhere.

In multicomponent systems, there are two mechanisms for mass
transport: convection and diffusion.  Convection is the flow of the
overall fluid, while diffusion occurs where the average velocities of
components are different.  The velocity of the overall fluid is a
macroscopic quantity because it is conserved, but the average
velocities of the components are not.  The macroscopic velocity of the
fluid $\mathbf{u}$ can be expressed in terms of the density $\rho^\sigma$ and
velocity $u^\sigma$ of each component in the form of
\begin{equation} 
 {\bf u} \equiv \frac{ \sum_\sigma  \rho^\sigma {\bf u}^\sigma}{ \sum_\sigma \rho^\sigma}. \label{uaverage} 
\end{equation} 
With the notation
\begin{equation} 
\Delta {\bf u}^\sigma \equiv {\bf u}^\sigma - {\bf u}, \label{dsl}
\end{equation}
the flux of each component can be divided into a convection part ${\bf
J}^{\sigma c}$ and a diffusion part ${\bf J}^{\sigma d}$:
\begin{equation}
 {\bf J}^\sigma \equiv \rho^\sigma {\bf u}^\sigma = \rho^\sigma ( {\bf u} +  \Delta {\bf u}^\sigma )
 = {\bf J}^{\sigma c} + {\bf J}^{\sigma d}.  \label{jjj}
\end{equation}  
Because mass conservation still holds for each component, the
continuity equation for each component is valid:
\begin{equation} 
\partial_t \rho^\sigma + \nabla \cdot {\bf J}^\sigma = 0 . \label{onecon}
\end{equation} 
Substituting Eq.~(\ref{jjj}) into Eq.~(\ref{onecon}), the convection
diffusion equation for a component can be obtained.
\begin{equation} 
\partial_t \rho^\sigma + \nabla \cdot {\bf J}^{\sigma c} = - \nabla \cdot {\bf J}^{\sigma d} .  \label{condieq}
\end{equation} 
From Eqs.~(\ref{uaverage}) and (\ref{dsl}), we see that 
\begin{equation} 
 \sum_\sigma {\bf J}^{\sigma d} = 0,   \label{abab}
\end{equation}  
which ensures the recovery of the continuity equation for the overall
system.  The diffusion process between two components is related to
the difference of the chemical potential of the two components, which
is also called the exchange chemical potential \cite{jones}.
Recognizing that the gradient of the exchange chemical potential
determines the diffusion processes, we obtain a first order
approximation for the diffusion flux of one component into all other
components as
\begin{equation} 
{\bf J}^{\sigma d} = - \sum_{\sigma '} M^{\sigma \sigma '} \nabla ( \mu^\sigma - \mu^{\sigma'}) ,
\label{akak}
\end{equation} 
where $\sigma$ and $\sigma '$ enumerate the components; $\mu^\sigma$ and $\mu^{\sigma '}$ are
the chemical potentials of components $\sigma$ and $\sigma '$; and $M^{\sigma \sigma '}$
is a symmetric positive definite mobility tensor.

A simple model for the diffusion process assumes that a diffusion flux
between two components is proportional to the overall density and the
concentration of each component.  Then mobility tensor can be expressed as
\begin{equation} 
M^{\sigma \sigma '}  = k^{\sigma \sigma '} \frac{\rho^\sigma \rho^{\sigma '}}{\rho} ,
\label{mss}
\end{equation}
where $k^{\sigma \sigma '}$ is the constant diffusion coefficient between
components $\sigma$ and $\sigma '$. It depends on components but is independent
of the total densities and concentration of each component.
Substituting Eq.~(\ref{mss}) into Eq.~(\ref{akak}), we have 
\begin{equation} 
 {\bf J}^{\sigma d} =   - \sum_{\sigma '} k^{\sigma \sigma '} \frac{\rho^\sigma \rho^{\sigma '}}{\rho} \nabla ( \mu^\sigma - \mu^{\sigma'}).
\label{bfjj}
\end{equation} 
Substituting Eq.~(\ref{bfjj}) into Eq.~(\ref{condieq}), the general
form of a convection diffusion equation is obtained as
\begin{equation} 
\partial_t \rho^\sigma + \nabla (\rho^\sigma {\bf u}) =
  \nabla  \sum_{\sigma '} k^{\sigma \sigma '} \frac{\rho^\sigma \rho^{\sigma '}}{\rho}
 \nabla ( \mu^\sigma - \mu^{\sigma '}) .
\label{gfcd}
\end{equation}

\section{Lattice Boltzmann for Multicomponent System}
To simulate a multicomponent fluid using LB we set up a LB equation
for each component.  The LBE for a component $\sigma$ of a multicomponent
system is given by
\begin{eqnarray} 
&& f^\sigma_i ({\bf r} + {\bf v_i} \Delta t , t + \Delta t) - f^\sigma_i ({\bf r}, t)\nonumber \\
 &=& \Delta t  \left[ \frac{1}{\tau} 
 \Big(f^{\sigma e}_i ({\bf r}, t) - f^{\sigma}_i ({\bf r},t) \Big) + F^\sigma_i \right], \label{ssLB}
\end{eqnarray} 
where $f^\sigma_i ({\bf r},t )$ is the particle distribution function with
velocity ${\bf v}_i$ for component $\sigma$, $f^{\sigma e}({\bf r},t)$ is its
equilibrium distribution and $F^\sigma_i$ is the forcing term of component $\sigma$
due to the mean potential field generated by the interaction of the
component $\sigma$ with the other components. The main task in setting up
this lattice Boltzmann method is to determine the correct form of the
forcing term $F^\sigma_i$ which will recover the convection diffusion equation
(\ref{gfcd}).

The density of each component and the total density are given by
\begin{eqnarray} 
\rho^\sigma &=& \sum_i f_i^\sigma, \\
\rho &=& \sum_\sigma \rho^\sigma.
\end{eqnarray}   
The average velocity of one component $\sigma$ and the overall fluid can be
defined as
\begin{eqnarray} 
 \rho^\sigma u^\sigma_\alpha  & \equiv & \sum_i f^\sigma v_{i\alpha} ,\\
 \rho u_\alpha & \equiv & \sum_\sigma \rho^\sigma u_{\alpha}^\sigma ,
\end{eqnarray} 
where $u^\sigma_\alpha $ is the average velocity of the component $\sigma $, and $u_\alpha
$ is the average velocity of the overall fluid.

The moments of equilibrium distributions for one component are chosen
to be
\begin{eqnarray} 
\sum_i f^{\sigma e}_i &=& \rho^\sigma   \nonumber, \\
\sum_i f^{\sigma e}_i  v_{i \alpha} & =  & \rho^\sigma u_\alpha, \nonumber \\
\sum_i f^{\sigma e} v_{i\alpha} v_{i\beta} & = &\frac{\rho^\sigma}{3} \delta_{\alpha \beta}
 + \rho^\sigma u_\alpha u_\beta, \nonumber\\
\sum_i f^{\sigma e} v_{i\alpha} v_{i\beta} v_{i\gamma} & = & 
\frac{\rho^\sigma}{3}(u_\alpha \delta_{\alpha \beta}  + u_\beta \delta_{\alpha \gamma} 
+ u_\gamma \delta_{\alpha \beta} )
\end{eqnarray} 
The moments for the forcing terms of one component are
\begin{eqnarray} 
\sum_i F^\sigma_i &=& 0  \label{ff1} \\
 \sum_i F^\sigma_i v_{i\alpha} &=& \rho^\sigma a^\sigma_\alpha, \label{ff2}\\
 \sum_i F^\sigma_i v_{i\alpha} v_{i\beta} &=& \rho^\sigma (a^\sigma_\alpha u^\sigma_\beta 
+ a^\sigma_\beta u^\sigma_\alpha),  \label{ff3} \\
 \sum_i F^\sigma_i v_{i\alpha} v_{i\beta} v_{i\gamma} &=& \frac{1}{3} \rho^\sigma   
 ( a^\sigma_\alpha \delta_{\beta \gamma} +  a^\sigma_\beta \delta_{\alpha \beta}  +  a^\sigma_\gamma \delta_{\alpha \beta} ) .
\end{eqnarray}
To utilize the analysis of the one component system we can establish a
LB equation for the total density by defining
\begin{eqnarray} 
\sum_\sigma f_i^\sigma &=& f_i, \nonumber\\
\sum_\sigma F_i^\sigma &=& F_i, \nonumber\\
\sum_\sigma \rho^\sigma a^\sigma_\alpha &=& \rho a_\alpha .    
\end{eqnarray} 
Similar to the counterparts of the one-component system, the moments
for the overall equilibrium distribution function are given by
\begin{eqnarray}
\sum_i f^e_i  &=& \rho ,\nonumber\\
 \sum_i f^e_i v_{i\alpha} &=& \rho u_\alpha , \nonumber \\
\sum_i f_i^e v_{i\alpha}v_{i\beta} &=&  \frac{1}{3} \rho \delta_{\alpha \beta} + \rho u_\alpha  u_\beta , \nonumber\\ 
\sum_i f_i^e v_{i\alpha} v_{i\beta} v_{i\gamma} &=& \frac{1}{3} \rho (u_\alpha \delta_{\beta \gamma} + u_\beta \delta_{\alpha \gamma} 
 + u_\gamma \delta_{\alpha \beta}) \nonumber \\
  &&  + \rho u_\alpha u_\beta u_\gamma + Q_{\alpha \beta \gamma}. 
\end{eqnarray}
The moments for the overall force terms are then given by
\begin{eqnarray}
 \sum_i F_i &=& 0 , \nonumber \\
 \sum_i F_i v_{i\alpha} &=& \rho a_\alpha, \nonumber 
\end{eqnarray}
Using Eq.~(\ref{ff3}), we obtain
\begin{eqnarray} 
 && \sum_i F_i v_{i\alpha} v_{i\beta} \nonumber \\
 &=& \sum_\sigma ( a_{\alpha}^\sigma u_{\beta}^\sigma + a_{\beta}u_{\alpha}^\sigma) \nonumber \\
 &=& \rho (a_\alpha u_\beta + a_\beta u_\alpha) + \sum_\sigma ( a_\alpha^\sigma \Delta u_{\beta}^\sigma 
                + a_{\beta}^\sigma \Delta u_{\alpha}^\sigma),  \label{olso}
\end{eqnarray} 
where the second term of Eq.~(\ref{olso}) is of a higher order
smallness than the first terms, and therefore does not enter the
hydrodynamic equations to second order.  For the third moment we have
\begin{equation}   
 \sum_i F_i v_{i\alpha} v_{i\beta} v_{i\gamma} = \frac{1}{3} \rho   
 (a_\alpha \delta_{\beta \gamma} + a_\beta \delta_{\alpha \beta}  +  a_\gamma \delta_{\alpha \beta} ) .
\end{equation}
By summing Eq.~(\ref{ssLB}) over $\sigma$, an effective LB equation for the
total density is
\begin{eqnarray} 
&&f_i ({\bf r} + {\bf v}_i \Delta t , t + \Delta t) - f_i ({\bf r}, t)  \nonumber \\
&=& \Delta t  \left[ \frac{1}{\tau} \Big(f^e_i ({\bf r}, t) - f_i ({\bf r},t) \Big) + F_i \right] , \label{sgLB}
\end{eqnarray} 
This is identical to the LB equation for a system of one component.
Therefore, the continuity equation and the Navier Stokes equation of
the overall fluid of a multicomponent system are recovered as
\begin{equation} 
\partial_t \rho + \partial_\alpha ( \rho U_\alpha ) = 0 + O (\epsilon^3),  \label{scon}
\end{equation}
where $U_\alpha \equiv u_\alpha + a_\alpha \Delta t / 2$ is the macroscopic velocity of the
fluid.  The Navier Stokes equation for the overall fluid is:
\begin{eqnarray} 
&& \partial_t (\rho U_\beta) + \partial_\alpha ( \rho U_\alpha U_\beta)  \nonumber \\
&=& - \partial_\alpha \left( \frac{1}{3} \rho \delta_{\alpha \beta} \right)  \nonumber \\
   &&+  \partial_\alpha \left( \frac{w}{3}  \rho (\partial_\alpha U_\beta + \partial_\beta U_\alpha) \right) + \rho a_\beta  \nonumber \\ 
& & - w \partial_\gamma  \partial_\alpha \rho u_\alpha u_\beta u_\gamma .
\label{snse}   
\end{eqnarray}
To recover the convection diffusion equation of each component, we
performed a Taylor expansion on the left of Eq.~(\ref{ssLB}) to second
order:
\begin{eqnarray} 
&& \Delta t (\partial_t + v_{i\alpha} \partial_\alpha) f^\sigma_i 
 + \frac{(\Delta t )^2}{2}  (\partial_t + v_{i\alpha} \partial_\alpha)^2  f^\sigma_i
 + O(\epsilon^3)  \nonumber  \\
&=&  \Delta t \left( \frac{1}{\tau} (f^{\sigma e}_i - f^\sigma_i ) +F^\sigma_i \right)   .      \label{s2LB}
\end{eqnarray} 
Because of the recursive nature of Eq.~(\ref{s2LB}), $f^\sigma_i$ can be
expressed by $f_i^{\sigma e }$ and derivatives of $f^{\sigma e}_i$ as
\begin{eqnarray} 
f^\sigma_i &=& f_i^{\sigma e}  + \tau F^\sigma_i \nonumber \\
            &&- \tau (\partial_t + v_{i\alpha}\partial_\alpha )
 (f_i^{\sigma e} + \tau F^\sigma_i) + O(\epsilon^2) .
\label{bfe}
\end{eqnarray} 
Substituting Eq.~(\ref{bfe}) into the left side of
Eq.~(\ref{s2LB}) we obtain
\begin{eqnarray} 
&&(\partial_t + v_{i\alpha} \partial_\alpha) (f^{\sigma e}_i  + \tau F^\sigma_i) \nonumber \\
 &-& w (\partial_t + v_{i\alpha}  \partial_\alpha)^2 (f^{\sigma e}_i  + \tau F^\sigma_i)
 + O(\epsilon^3) .  \nonumber \\
&=&   \frac{1}{\tau} (f^{\sigma e}_i +  \tau F^\sigma_i - f^\sigma_i ).         \label{s3LB}
\end{eqnarray} 
Summing Eq.~(\ref{s3LB}) over $i$ gives,
\begin{eqnarray} 
&& \partial_t \rho^\sigma + \partial_\alpha (\rho^\sigma u_\alpha) + \tau \partial_\alpha ( \rho^\sigma a^\sigma_\alpha) \nonumber\\
 &-& w \sum_i (\partial_t + v_{i\alpha}  \partial_\alpha)^2 (f^{\sigma e}_i  + \tau F^\sigma_i) =  O(\epsilon^3)  
\label{sgao}
\end{eqnarray}
The first moment of $f_i^e$ and $f_i^{\sigma e}$ are not identical,
and the continuity equation cannot be obtained.  Eq.~(\ref{sgao})
shows that $\partial_t \rho^\sigma + \partial_\alpha (\rho^\sigma u_\alpha) + \tau \partial_\alpha ( \rho^\sigma a^\sigma_\alpha)$ is of order
$O(\epsilon^2)$, and $F^\sigma_i $ is of order $O(\epsilon)$.  Therefore $\partial_t \rho^\sigma + \partial_\alpha
(\rho^\sigma u_\alpha) $ is of order $O (\epsilon^2)$, and we get
 \begin{eqnarray*}
&& w \sum_i (\partial_t + v_{i\alpha}  \partial_\alpha)^2 (f^{\sigma e}_i  + \tau F^\sigma_i)\\
&=& w \partial_\beta \left[ \partial_t (\rho^\sigma u_\beta ) + \partial_\alpha (\frac{\rho^\sigma}{3} 
 + \rho^\sigma u_\alpha u_\beta )  \right] + O(\epsilon^3) .
\end{eqnarray*}
So Eq.~(\ref{sgao}) can be simplified to
\begin{eqnarray}  
 && \partial_t \rho^\sigma  + \partial_\alpha (\rho^\sigma u_\alpha) + \tau  \partial_\alpha  \rho^\sigma a^\sigma_\alpha + O(\epsilon^3)\nonumber \\
&&- w \partial_\beta \left[ \partial_t (\rho^\sigma u_\beta) 
 + \partial_\alpha (\frac{\rho^\sigma}{3}  
 + \rho^\sigma u_\alpha u_\beta ) \right] = 0 .  \label{ssim}
\end{eqnarray} 
Eqs.~(\ref{snse}) yields
\begin{equation} 
\partial_t U_\beta = - U_\alpha \partial_\alpha U_\beta - \frac{1}{\rho} 
 \partial_\alpha \left( \frac{\rho}{3} \delta_{\alpha \beta} 
  \right) + a_\beta + O(\epsilon^2) .
\label{stu}
\end{equation}  
From Eq.~(\ref{sgao}) it follows that
\begin{equation} 
\partial_t \rho^\sigma = - \partial_\alpha ( \rho^\sigma U_\alpha ) + O(\epsilon^2).
\label{sxx}
\end{equation} 
Inserting Eqs.~(\ref{stu}) and (\ref{sxx}) into Eq.~(\ref{ssim}) we get
\begin{eqnarray} 
\partial_t ( \rho^\sigma u_\beta )  =  &-& \partial_\alpha (\rho^\sigma U_\alpha U_\beta) 
 - \frac{\rho^\sigma}{\rho} \partial_\alpha \left( \frac{\rho}{3} \delta_{\alpha \beta} \right) \nonumber \\
&+&\rho^\sigma a_\beta  + O(\epsilon^2) .
\label{jeep}
\end{eqnarray} 
Substituting Eq.~(\ref{jeep}) into Eq.~(\ref{ssim}) results in
\begin{eqnarray} 
\partial_t \rho^\sigma  + \partial_\alpha (\rho^\sigma U_\alpha)
&=& \partial_\alpha \bigg[ \tau \rho^\sigma a_\alpha- \tau \rho^\sigma a^\sigma_\alpha  
    + w \partial_\alpha  \left( \frac{\rho^\sigma}{3} \right)
\nonumber \\&&
- w \frac{\rho^\sigma}{\rho} \partial_\alpha \left( \frac{\rho}{3} \right) \bigg]  , 
\label{geq}
\end{eqnarray} 
From this we deduce that the correct form of the forcing term is
\begin{eqnarray} 
\rho^\sigma a_\alpha^\sigma &\equiv& - \frac{w}{\tau} \rho^\sigma \partial_\alpha ( \mu^\sigma - \frac{1}{3} \ln \rho^\sigma ) \nonumber \\
&=& - \frac{w}{\tau} \left( \rho^\sigma \partial_\alpha \mu^\sigma - \frac{1}{3}  \partial_\alpha \rho^\sigma \right) ,
\label{definef} 
\end{eqnarray} 
where the coefficient is $\frac{w}{\tau} = 1- \frac{\Delta t}{2 \tau} $.  This
coefficient approaches 1 as $\Delta t$ approaches 0, as one would expect
from the continuum limit. This constitutes the main result of this
paper.  Plugging Eq.~(\ref{definef}) into Eq.~(\ref{geq}), we then
obtain the convection diffusion equation
\begin{eqnarray} 
\partial_t \rho^\sigma  + \partial_\alpha (\rho^\sigma U_\alpha )  
&=& \partial_\alpha \left[ w \sum_{\sigma'} \frac{\rho^\sigma\rho^{\sigma'}}{\rho}  \partial_\alpha ( \mu^\sigma - \mu^{\sigma'}) \right] . 
\label{woche}
\end{eqnarray}

The diffusion flux of component $\sigma$ is
\begin{equation} 
J^{\sigma d} = - w  \sum_{\sigma'} \frac{\rho^\sigma\rho^{\sigma'}}{\rho} \partial_\alpha ( \mu^\sigma - \mu^{\sigma'}).
\label{gogogo}
\end{equation} 
So that the $w$ in Eq.~(\ref{gogogo}) is equivalent to $k^{\sigma \sigma'}$ in
Eq.~(\ref{bfjj}).

\section{Numerical validation}
We examined the equilibrium behavior of phase separated binary and
ternary systems. We used the Flory-Huggins free which is a very
popular model to study polymer solutions. It is given by
\begin{eqnarray} 
 \mathcal{F}  &=& \int \left( - \sum_\sigma \theta n^\sigma m^\sigma 
 + \sum_\sigma  n^\sigma \theta \ln \phi^\sigma   \right. \nonumber \\
&& \qquad + \left. \sum_\sigma\sum_{\sigma'} \frac{1}{2} \chi^{\sigma \sigma'} \theta m^\sigma n^\sigma
 \phi^{\sigma'}  \right) dV ,
\label{lgf}
\end{eqnarray}  
where $m^\sigma$ is the polymerization of the component $\sigma$, $n^\sigma$ is its
number density, and $\phi^\sigma$ is its volume fraction. It is defined as
\begin{equation} 
\phi^\sigma = \frac{m^\sigma n^\sigma}{\sum_\sigma ( m^\sigma n^\sigma)} = \frac{\rho^\sigma}{\rho}, 
\end{equation}   
where $ \rho^\sigma $ is the mer density of component $\sigma$ and $ \rho$ is the mer
density of the system, which is a constant in the Flory-Huggins model.
To validate our algorithm we compared the binodal lines obtained by
our algorithm to the theoretical ones obtained by minimizing the free
energy.  We used the interfacial tension parameter $\kappa =0$ in all our
LB simulations of binary and ternary systems, because there is an
intrinsic surface tension in the LB simulation due to higher order
terms \cite{wagner2006a}, which did not appear explicitly in the
second order Taylor expansion presented in this paper. Since we are
only evaluating the phase-behavior here we use a one dimensional model
known as D1Q3. This model has the velocity set $\{v_i\} =
\{-1,0,1\}$. This is an important test since all other frequently used
higher dimensional model have this D1Q3 model as a projection.

We consider two binary systems: a monomer system with $m^A = 1$ and
$m^B=1$, and a polymer system with $m^A = 10$ and $m^B =1$.  For both
systems, the total density was $\rho = 100$.  Throughout this paper we
choose the self interaction parameters to vanish: $\chi^{\sigma\sigma}=0$. The
critical volume fractions for the monomer system are $\phi^A =0.5$ and
$\phi^B = 0.5$ and for the polymer systems are $\phi^A =0.24$ and $\phi^B =
0.76$.  To induce phase separation a small sinusoidal perturbation
$\Delta\phi(x)$ was added in the initial conditions.  The amplitude of the
perturbation is 0.1 and its wavelength is the lattice size.  The
initial volume fraction of component A is given by $\phi^A (x) = \phi^{A0} +
\Delta \phi (x)$.  The initial volume fraction of component B is given by $\phi^B
(x) = \phi^{B0} - \Delta \phi (x)$.

\begin{figure}
\begin{center}
\resizebox{\columnwidth}{!}{\rotatebox{0}{\includegraphics{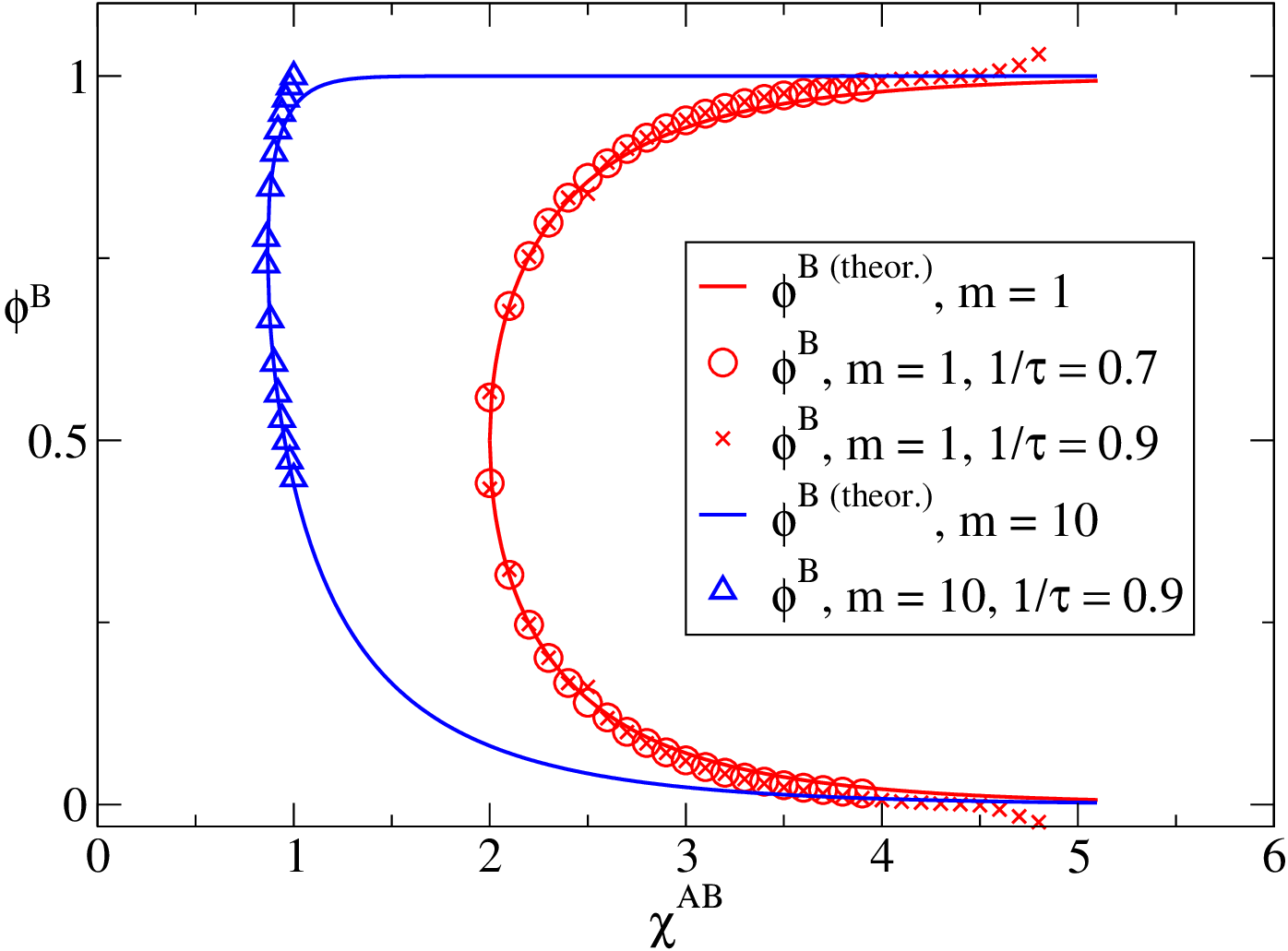}}}
\caption{(Color online) Comparison of the binodal lines for monomer
  and polymer systems show good agreement with theory. For $m^A=1,
  m^B=1$ we also show that the results are independent of the
  relaxation time.  For the polymer system with $m^A = 10$, $m^B=1$, the
  binodal points by LB with $1/\tau = 0.9$ the match is slightly less
  good for large volume fractions of $\phi^B$. }
\label{binary_binodal}  
\end{center}
\end{figure}   
\begin{figure}
\begin{center}
\resizebox{0.8\columnwidth}{!}{\rotatebox{0}{\includegraphics{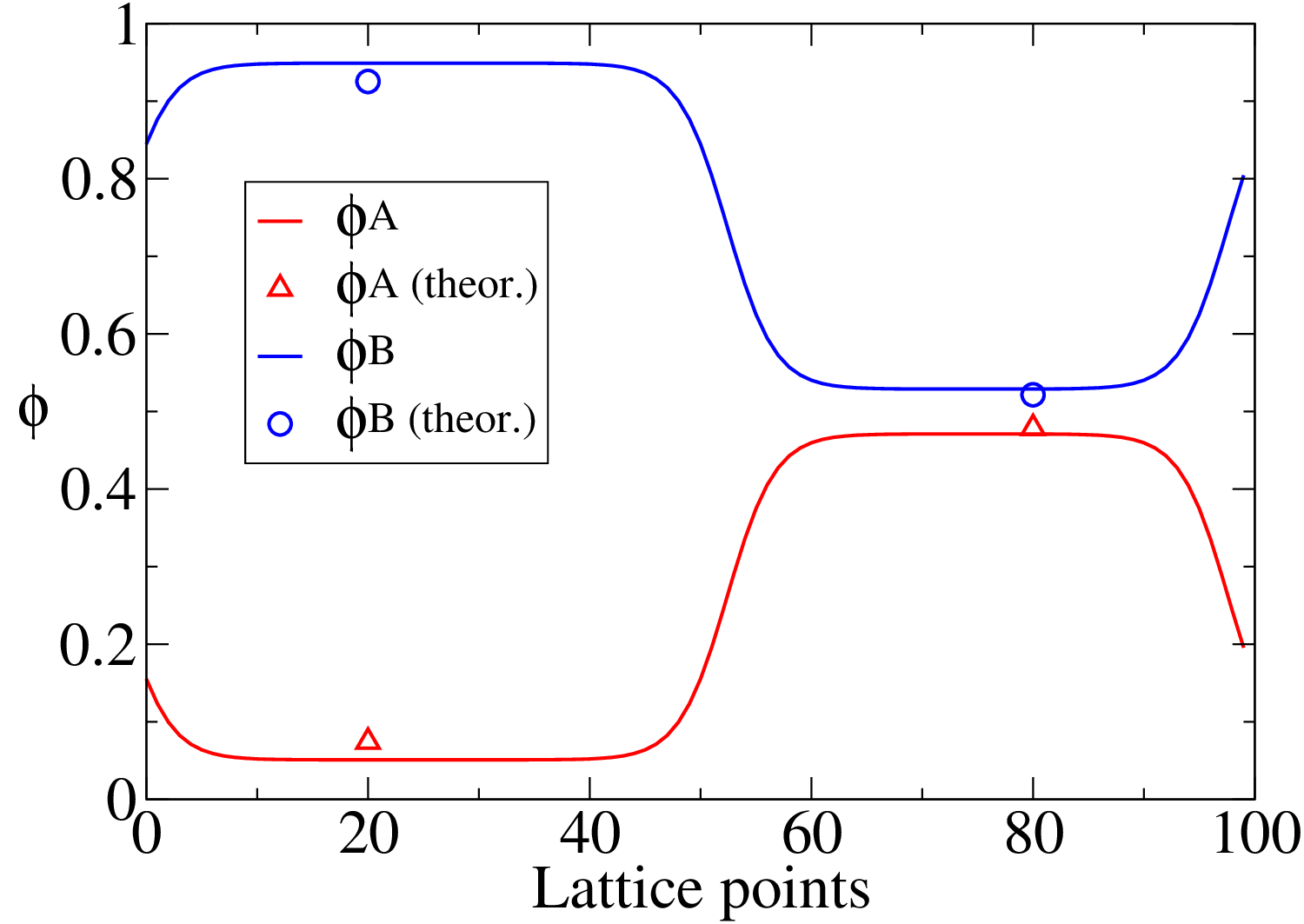}}}
\resizebox{0.8\columnwidth}{!}{\rotatebox{0}{\includegraphics{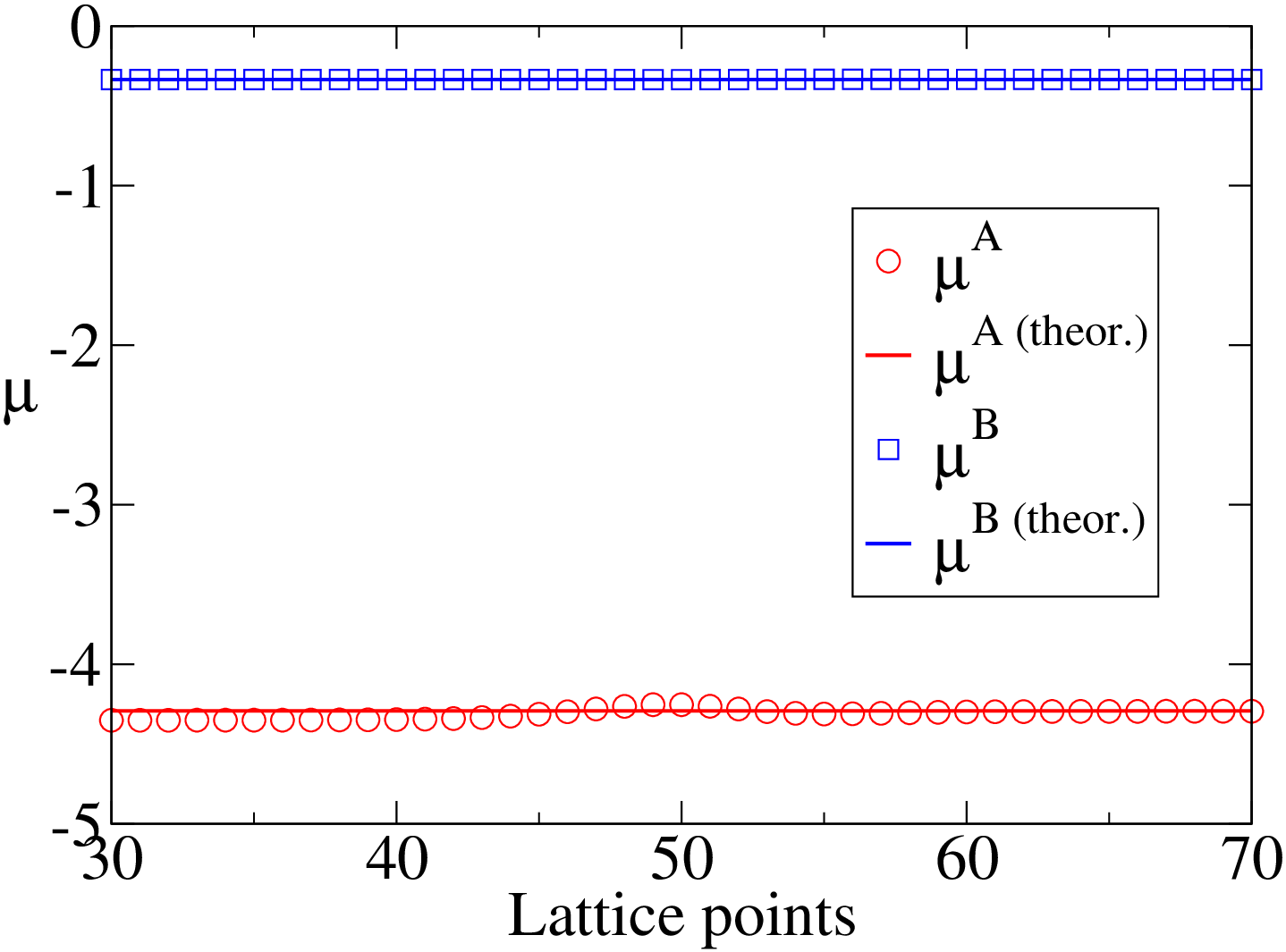}}} 
\caption{(Color online) The volume fraction and the chemical potentials of the two
components for the polymer-monomer mixture.  The parameters are $m^A$ =
10, $m^B$ = 1, $\chi^{AB}= 0.94$, $\kappa = 0$, and $1/ \tau = 0.9 $. }
\label{LBchem}
\end{center}  
\end{figure}

The monomer system was simulated with different inverse relaxation
times. In Figure \ref{binary_binodal} we show results for $ 1/\tau =
0.7$, and $0.9$. We see that the equilibrium densities have only a
very slight dependence on the relaxation time, although the range of
stability depends noticeably on the relaxation time.  The polymer
system was simulated with only one inverse relaxation time of $ 1/ \tau =
0.9$.  Starting from the critical point and increasing the $\chi^{AB}$ value
for each initial condition until the simulations were numerically
unstable, we obtained a pair of binodal points for each initial
condition.  The system reached a stable state after about 5000 time
steps.  The measurement were taken after 50000 time steps to be sure
that an equilibrium state had been reached.

For the polymer system, Figure~\ref{LBchem} shows the comparison of
the total density, and the volume fractions and chemical potentials of
each component to the corresponding theoretical values.  The total
density of a system in equilibrium by LB is essentially constant with
a variation of $\Delta\rho/\rho<10^{-5}$. The volume fractions of each component
in the LB simulation agree well with the theoretical values.
The chemical potential of each component by the LB simulation was very
close to the theoretical value.  The chemical potential $\mu_A$,
corresponding to the polymer component, varied slightly with a
difference for the bulk values of about $2\cdot10^{-2}$ and a variation in
the interface of about $4\cdot10^{-2}$. This is the
underlying reason for the small deviation from the theoretically
predicted concentration. The potential $\mu_B$ was nearly constant with
a variation of less than $10^{-4}$ in the bulk and a variation of about
$10^{-3}$ at the interface. For large values of $\chi^{AB}$ this
discrepancy increases leading to the noticeable variation of the
equilibrium densities of the polymer system as shown in Figure
\ref{binary_binodal}.

\begin{figure}
\begin{center}
\resizebox{0.9\columnwidth}{!}{\rotatebox{0}{\includegraphics{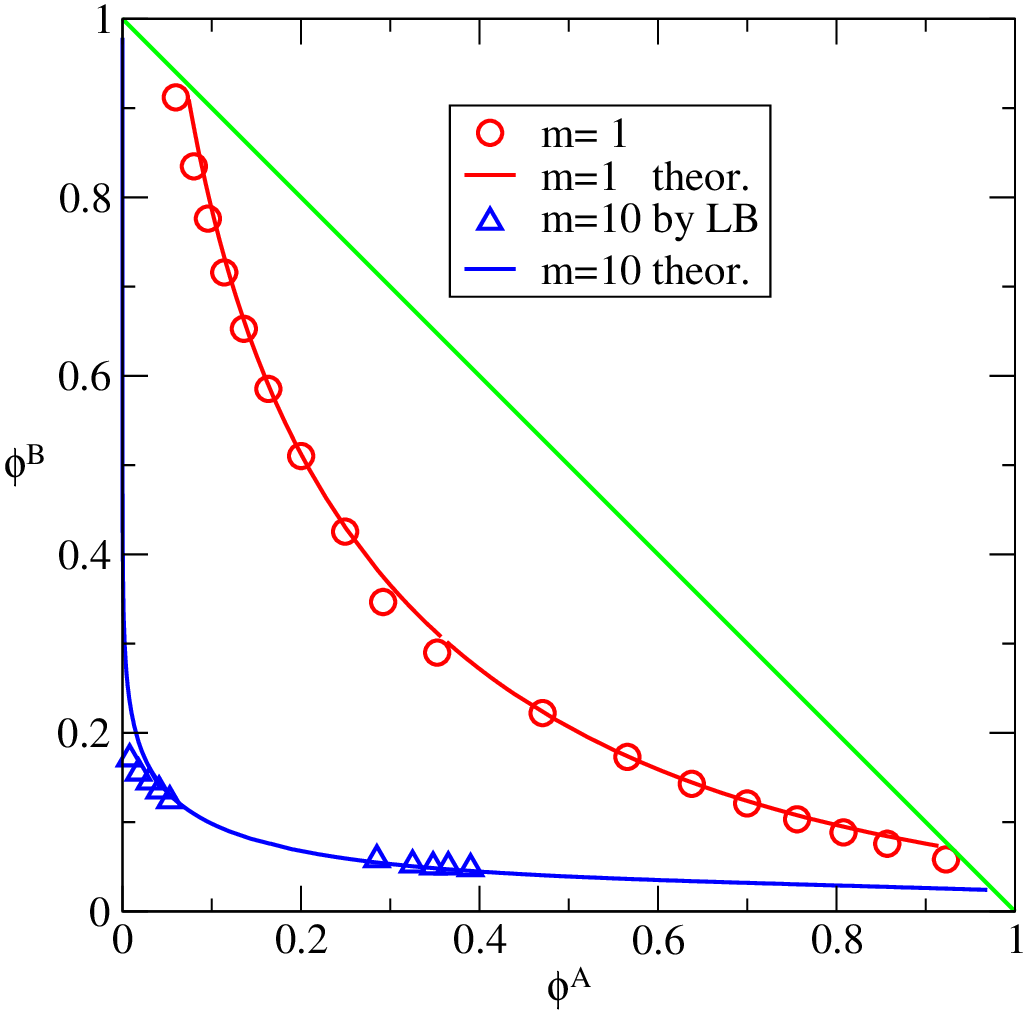}}}
\caption{(Color online) The binodal points of two ternary systems
obtained by LB simulation. The monomer system has $m^A = 1$, $m^B=1$,
and $m^C=1$; the polymer system has $m^A = 10$, $m^B=1$, and $m^C=1$.
The parameters for both systems are $\chi^{AB}$ =3, $\chi^{AC}$ = 0.5,
$\chi^{BC}$ = 0.2 and $ 1/\tau = 0.9$.}
\label{triLBm10phase}  
\end{center}
\end{figure}
\begin{figure}
\begin{center}
\resizebox{0.8\columnwidth}{!}{\includegraphics{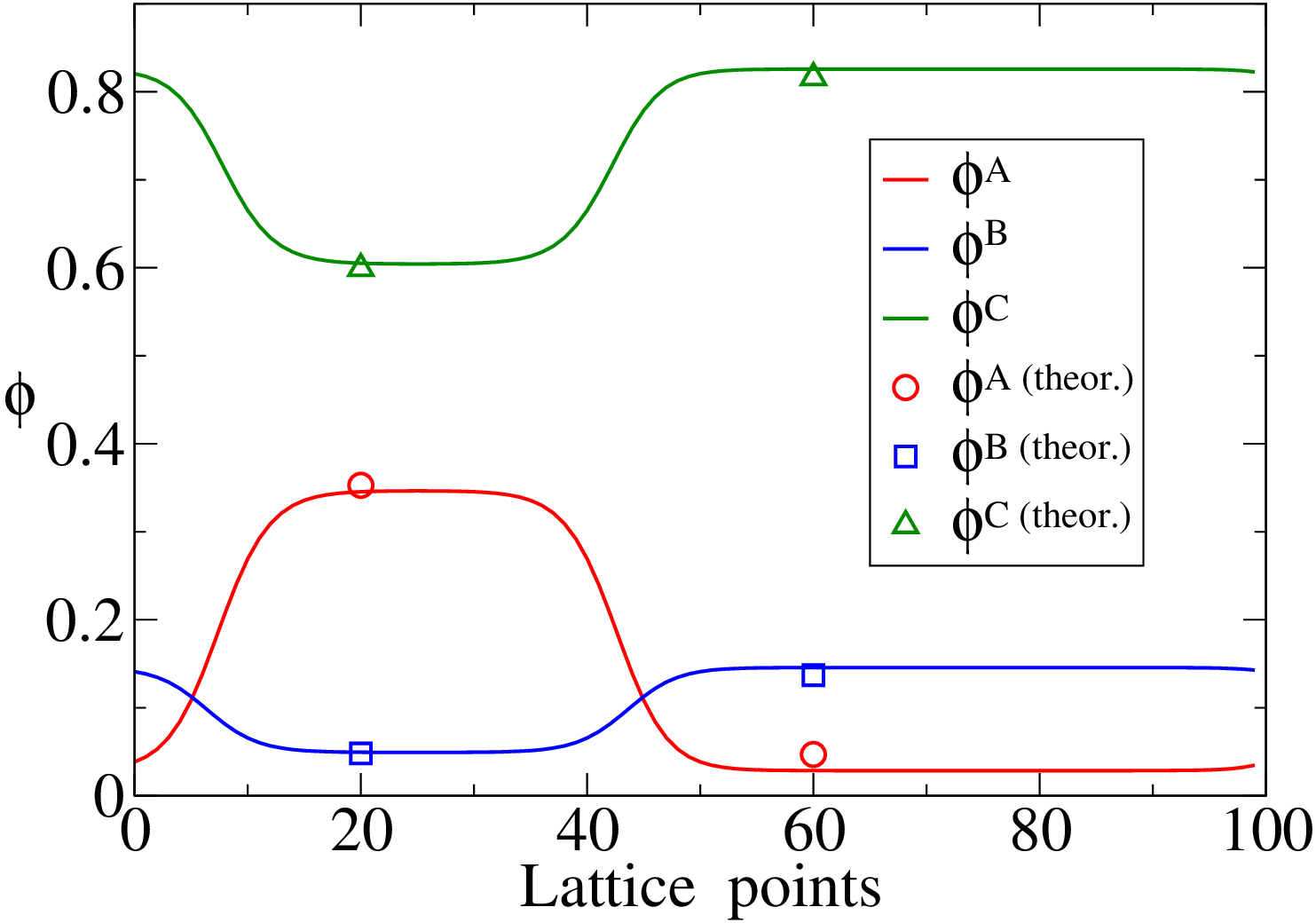}}
\resizebox{0.8\columnwidth}{!}{\includegraphics{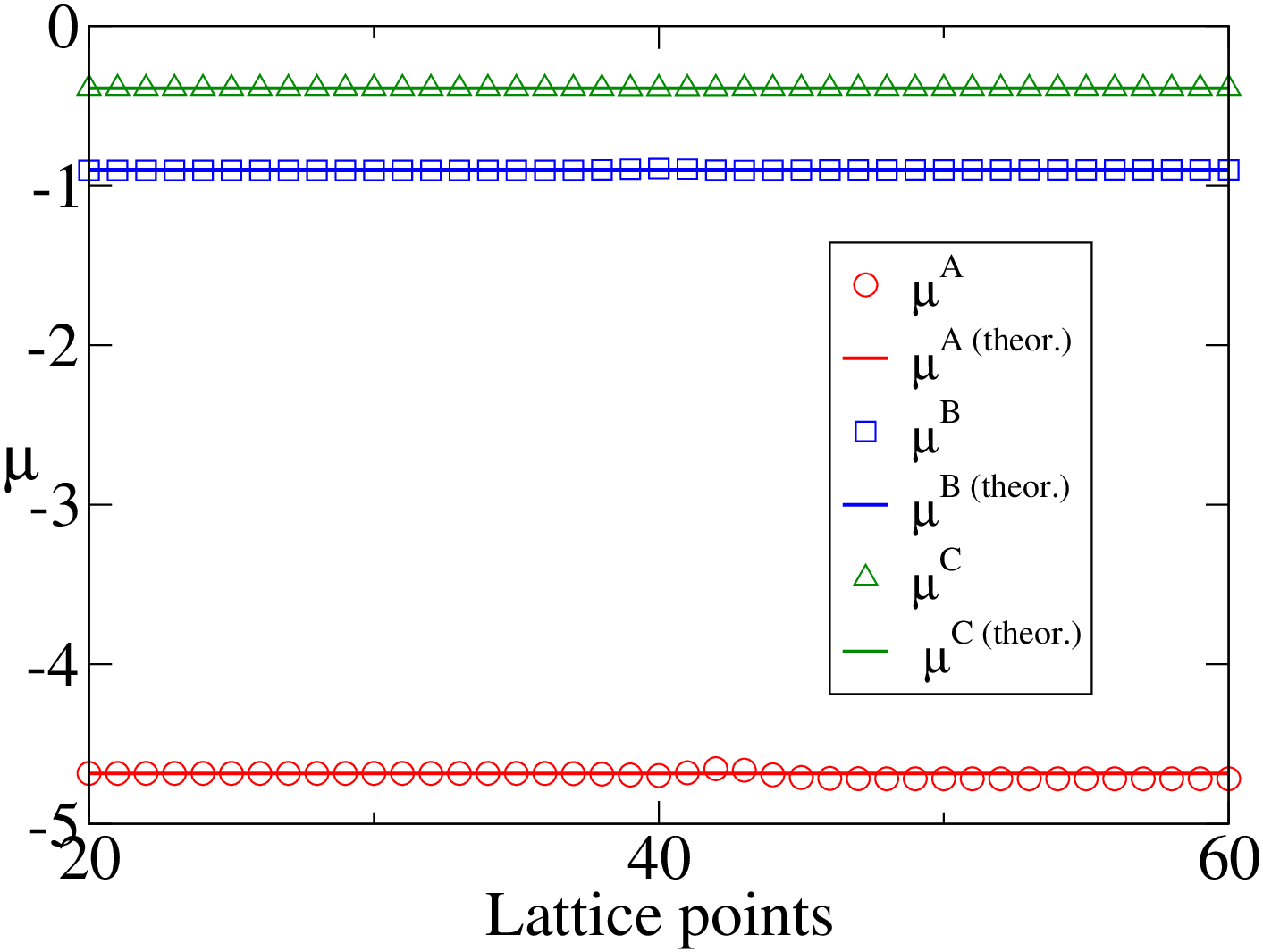}}
\caption{(Color online) The volume fractions and the chemical
  potentials for three component polymer system. The parameters are
  $m^A = 10$, $m^B = 1$, $m^C = 1$,
  $\chi^{AB}=3$, $\chi^{AC} = 0.5$, $\chi^{BC} = 0.2$, and $1/\tau = 0.9$. The
  initial homogeneous volume fractions were $\phi_A=0.14$, $\phi_B=0.11$,
  and $\phi_C=0.75$.}
\label{triLBm10ch}
\end{center}  
\end{figure}

We also performed LB simulations with two ternary systems: a monomer system
with $m^A = 1$, $m^B=1$, and $m^C=1$ and a polymer system with $m^A =
10$, $m^B =1$, and $m^C =1$.  The $\chi$ parameters for both systems were
$\chi^{AB} =3$, $\chi^{AC} = 0.5$, and $\chi^{BC} = 0.2$.  The other $\chi$
parameters were zero. The inverse relaxation time constant for both
simulations was $1/ \tau = 0.9$.  The critical point for the monomer
system was $\phi^{A,cr} = 0.32$, $\phi^{B,cr} = 0.32$, and $\phi^{C,cr} = 0.36$.  The critical
point for the polymer system was $\phi^{A,cr} = 0.14$, $\phi^{B,cr} = 0.11$, and $\phi^{C,cr}
=0.75$.  The initial state of each simulation were set from the
critical points towards the end point ($\phi^A = 0.5$, $\phi^B = 0.5$, $\phi^C
=0$).  Initially a small sinusoidal wave perturbation $\Delta \phi$ of an
amplitude of 0.1 and wavelength of the lattice size was superimposed
on the initial volume fraction of the A component. This perturbation
was subtracted from the B component and the C component was
constant. I performed a LB simulation for each set of initial volume
fractions and obtained the volume fractions of the two phases in the
equilibrium state, resulting in two binodal points.  The simulation
reached a stable state after about 20,000 time steps.  The
measurements were taken after 200,000 time steps to make sure the
equilibrium state was reached.

Figure~\ref{triLBm10phase} shows the comparison of the binodal
points by LB simulation to the theoretical binodal lines of both
systems.  The binodal points obtained by the LB simulation agree
fairly well with the theoretical binodal lines for the monomer and
polymer systems.  The simulation becomes unstable when $\phi^A$ is close
to zero, i.e. when one component is nearly depleted. In this region
the simulation results also deviate noticeably from the theoretical
binodal lines.  Immediately near the critical point, the evolution of
the system becomes extremely slow so the slight deviation between the
binodal points obtained through the LB simulation and the theoretical
ones probably indicates that the LB simulation was not yet fully
equilibrated.

For the polymer system Figure~\ref{triLBm10ch} shows a comparison of
the volume fractions and chemical potentials of each component.  The
total density of the system is again nearly constant with variation of
less than $\Delta\rho/\rho<10^{-4}$ in the bulk. At the interface there is a
small variation of $\Delta\rho/ \rho<10^{-2}$. The volume fractions of each phase
in the simulation were very close to their theoretical values. The
chemical potential of component A was slightly different in two phases
with a variation of about $2\cdot10^{-2}$, while the chemical potentials
of components B and C were much closer in the two phases with a
variation of less than $10^{-4}$.

\section{Outlook}
We have presented a general lattice Boltzmann algorithm for
systems with an arbitrary number of components which is based on an
underlying free energy. In this algorithm the key thermodynamic
quantities such as the chemical potentials of the components are
immediately accessible. It is also manifestly symmetric for all
components. We tested the equilibrium behavior of the new algorithm
for two and three component systems in each case examining both the
case of monomer and polymer mixtures with an underlying Flory-Huggins
free energy. We obtained to expected phase-diagrams to good accuracy
and the chemical potentials were constant to good accuracy for the
monomer systems. Polymer systems were more challenging to simulate but
still obtained acceptable results for $m=10$. Higher polymerizations,
however, become increasingly difficult to realize with the current
algorithm.

There are three directions in which we hope to extend this
algorithm in the future. The current algorithm does not allow for
component dependent mobility. We are working on developing an
algorithm that can recover an arbitrary mobility tensor
$\kappa^{\sigma\sigma'}$. The chemical potential is only approximately
constant. Recent progress for liquid-gas systems \cite{wagner2006a}
makes us hopeful that we will be able to ensure that the chemical
potential is constant up to machine accuracy. And lastly we hope to
extend to model so that it can simulate polymer systems with
significantly larger polymerization.

\end{document}